\documentclass[aps,prl,a4paper,twocolumn]{revtex4-1}
\usepackage{amsmath,amssymb,bm,graphicx,float}

\def \ket #1{{\vert #1\rangle}}
\def \bra #1{{\langle #1\vert}}

\def \k {\bm{k}}

\def \k {\mathbf{k}}
\def \r {\mathbf{r}}
\def \R {\bm{R}}

\newcommand {\q} {\mathbf{q}}

\def\bq{\mathbf{q}}

\def\bR{\mathbf{R}}

\def\Winf{W_\infty}

\def\zhat{\hat{\mathbf{z}}}

\usepackage{graphicx}
\begin{document} 
\title{Band geometry of fractional topological insulators}%
\author{Rahul Roy}
\affiliation{Department of Physics and Astronomy, University of California, Los Angeles}

\begin{abstract}
  Recent numerical simulations of flat band models with
  interactions which show clear evidence of fractionalized
  topological phases in the absence of a net magnetic field
  have generated a great deal of interest.  
  We provide an explanation for these observations by showing
  that the physics of these systems is the same as that of
  conventional fractional quantum Hall phases in the lowest
  Landau level under certain ideal conditions which can be
  specified in terms of the Berry curvature and the Fubini
  study metric of the topological band. In particular, we show
  that when these ideal conditions hold, the density operators
  projected to the topological band obey the celebrated
  $W_{\infty}$ algebra.  Our approach provides a quantitative
  way  of testing the
  suitability of topological bands for hosting fractionalized
  phases.

\end{abstract}

\maketitle

The advent of topological insulators which are band insulators
with topologically non-trivial bands, has generated a great
deal of recent interest in topological
phases~\cite{Hasan_10_3045,qi2011topological,hasan2011three}. The
Landau levels whose filling gives rise to the integer quantum
Hall effect~\cite{Klitzing_80_494} can also be regarded as
topologically non-trivial bands. While the integer quantum Hall
effect has so far only been observed in the presence of large
magnetic fields, a quantized Hall conductance can also arise in
the presence of a periodic potential, where it can be related
to a topological invariant associated with the bands of Bloch
wavefunctions~\cite{Thouless_82_405}.  In the absence of any
time-reversal symmetry breaking, this invariant has to be
zero. However, there exist tightbinding models which explicitly
demonstrate that a quantized Hall conductance is possible in a
net zero external magnetic field, which albeit break
time-reversal symmetry~\cite{Haldane_88_2015}.

In the presence of interactions, electrons in fractionally
filled Landau levels can form a liquid-like phase with a
quantized Hall conductance and a gap to all bulk
excitations~\cite{Stormer_83_1953}. This is a topological phase
with a non-trivial ground state degeneracy on a torus and
excitations with fractional charge and, depending on the
filling fraction, fractional or possibly even non-abelian
statistics. The question of whether similar phenomena can occur
in a band insulator model has recently been addressed in a
series of numerical works by many
groups~\cite{PhysRevLett.106.236804,sheng2011fractional,regnault2011fractional,2012arXiv1207.6094L,2012arXiv1207.3539S},
which provide clear evidence for the existence of gapped phases
possessing many of the signatures of the proposed ground states
for fractional quantum Hall states.

We provide a rationalization for these surprising developments
based on an approach introduced in
Ref.~\onlinecite{parameswaran2012fractional}.  We show that
under certain ideal conditions which will be specified in
detail below, the projected density operators obey a closed
algebra which has the same form as the celebrated $W_{\infty}$
algebra of the lowest Landau level projected density
operators~\cite{girvin1986magneto,girvin1985collective,cappelli1993infinite,karabali1994w}.

The single particle states of Chern bands are very different
from Landau level wavefunctions, so it is not a priori clear
why partially filled Chern bands can display an analog of the
fractional quantum Hall effect (FQHE). One set of
rationalizations for these numerical results has been based on
trial wavefunctions constructed either by mapping lowest Landau
level wavefunctions in the Landau gauge to Wannier
functions~\cite{qi2011generic} or by
partonization~\cite{PhysRevB.85.125105,PhysRevB.85.165134}. The
spectacular success of model wavefunctions such as Laughlin's
famous wavefunction in the theory of the FQHE~\cite{laughlin1983anomalous} makes such an
approach very attractive, but the lack of an appealing analytic
form of the Chern band wavefunctions and their poor overlap
with exact ground-state wavefunctions in small systems is in
sharp contrast to the model FQHE wavefunctions. This has
motivated the search for other explanations.

The essential requirements for the formation of fractional
Chern insulators were assumed, in the early numerical work, to
be some of the characteristic energetic features of the
FQHE. These are: 1. a nearly flat band with a non-trivial
topological invariant, and 2. short range interactions whose
energy scale is much larger than the band width of the
non-trivial band, but much smaller than the band gap.  Under
these circumstances, it is reasonable to project the
interactions to the topological band, as is usually carried out
in the theory of the FQHE. The assumption is that the low
energy spectrum consists of states whose admixture with
components from the other bands can be neglected. Even, with
this assumption, however, there are a large number of
parameters in the fractional Chern inuslator problem. First,
there is the freedom in choosing the lattice itself, which
breaks both continuous translational and rotational symmetry,
but may have certain discrete point and space group
symmetries. One may also vary the parameters that determine the
detailed form of the interaction and finally, one can also
change the various tight binding parameters of the single
particle Hamiltonian.

In numerical experiments to date, only a limited portion of
this large phase space of possible fractional Chern insulators
has been explored. Already, it is clear that there is a great
variation in the stability of fractional Chern insulator states
even when the basic energetic criteria listed in the previous
paragraph are met. One of the aims of this paper is to identify
other criteria which affect the stability of fractional Chern
insulator states. Our approach is based on studying the
commutation relations of projected density operators, a
direction which has yielded some success in analyzing
fractional Chern
insulators~\cite{2012arXiv1207.2133M,murthy2011composite,parameswaran2012fractional,2012EPJB...85...15G}.

Before delving into the role of interactions, we describe the
basic framework of (nearly) flat band topological
insulators. Our starting point is  a tightbinding
model which has topologically non-trivial bands, a famous
example of which is the Haldane model on a honeycomb
lattice~\cite{Haldane_88_2015}. By varying the tightbinding parameters, one can
flatten the energy bands without altering the topology of the
band structure~\cite{PhysRevLett.106.236802,PhysRevLett.106.236803}.  The Hamiltonian of an $N$ band
insulator can be written as $\sum_{\k,p,q} \ket{\k,p} (h(\k))_{pq}
\bra{\k,q} $, where the sum over crystal momenta is restricted
to the first Brillouin zone~(BZ). (Here and for the remainder of the
proposal, we will adopt the convention that repeated indices
are \textit{not} implicitly summed over. When required, we will
indicate a summation explicitly.)

The states  $\ket{\k,p}$ are the Fourier transforms of the
localized tightbinding orbital states:
\[
\ket{\k,p}
= \sum_{\R_n} e^{i \k. (\bm{e}_p+ \R_n)}
\ket{\R_n,p}\, , 
\] 
where $\bm{e}_p + \R_n$ denotes the position of the (localized)
$p^{\textrm{th}}$ orbital, $\ket{\R_n,p}$,  in the $n^{\textrm{th}}$ unit cell
situated at the lattice vector $\R_n$. The matrix $h(\k)$ can be diagonalized through an
appropriate unitary transformation and the Hamiltonian written
in the form $H_{K}=\sum_{\gamma,\k} E_{\gamma}(\k)
\ket{\k,\gamma}\bra{\k,\gamma}$ where $\ket{\k,\gamma} =
\sum_{p} u^{\gamma}_{p}(\k)\ket{\k,p} $ and $ (u^{\gamma}_j(\k))$
is a normalized eigenstate of $h(\k)$ with eigenvalue
$E_{\gamma}(\k)$. 

We will use the label $\alpha$ for the topological band of interest.
The Berry curvature, $B_{\alpha}(\k)$ of the band is defined as: 
\begin{align}
 B_{\alpha}(\k) = -i\sum_{p}\left(\frac{\partial {u}^{\alpha*}_p}{\partial
   k_x}\frac{\partial {u}^{\alpha}_p}{\partial k_y} -   
   \frac{\partial {u}^{\alpha*}_p}{\partial k_y}\frac{\partial {u}^{\alpha}_p}{\partial k_x}\right)  
 \end{align}
and its integral over the Brillouin zone is 
\begin{align}
  \int_{BZ} dk_x\, dk_y\,\, B_{\alpha}(\k) =  2\pi C_{\alpha}
\end{align}
where $C_\alpha$ is the Chern number of the band, $\alpha$. For a topological (Chern)
band, $C_\alpha$ is non-zero and without loss of generality, we can 
take $C_\alpha$ to be a positive integer.

We now consider the role of interactions by adding a term
$U_{\textrm{int}}$ to the Hamiltonian. The interactions we
consider are generally (but not always) density-density
interactions of the form
$U_{\textrm{int}}=\sum_{i,j}u(\r_i - \r_j)$. In the limit of a
large band gap, one can safely neglect the mixing between the
Chern band and the unfilled bands. If the bandwidth
is small compared to the scale of the interactions,
$E_{\alpha}(k)$ may be treated as constant and may be set to
zero by a simple regularization. 
With this approximation, the
low energy effective Hamiltonian including interactions has the
form: $H_{\textrm{eff}}=  \bar{U}_{\textrm{int}}$, where
$\bar{U}_{\textrm{int}}$ is the interaction projected to the Chern band.

One encounters a similar Hamiltonian in the treatment of
interactions in the lowest Landau level in a large magnetic
field. In that case, the effective Hamiltonian of a clean
system obtained by projecting density-density interactions to the lowest Landau
level has the form
\begin{align}
H_{LLL} = \frac{1}{2} \sum_{\bq} V(\bq)e^{-q^2\ell_B^2/4}
\overline{\rho}_\bq \overline{\rho}_{-\bq}  \label{LLL-proj-int}
\end{align}
where 
$\overline{\rho}_{\bq}$ differs from the projected density, 
$\mathcal{P} \rho_\bq \mathcal{P}$ by a $q$-dependent
constant, $ \mathcal{P} \rho_\bq \mathcal{P} =
e^{-q^2\ell_B^2/4} e^{i\bq\cdot\bR} \equiv e^{-q^2\ell_B^2/2}
\overline{\rho}_{\bq}$. 

In the LLL problem, the projected density operators, $\overline{\rho}_{\bq}$
obey the $\Winf$ algebra, first identified by Girvin, McDonald
and Platzman (GMP)~\cite{girvin1986magneto,girvin1985collective,cappelli1993infinite,karabali1994w}
: 
\begin{align}
[\overline{\rho}_{\bq_1},\overline{\rho}_{\bq_2}] =  2i \sin \left(\frac{\bq_1\wedge\bq_2\ell_B^2}{2}\right) \overline{\rho}_{\bq_1+\bq_2}
\label{GMP}
\end{align}
where $\bq_1\wedge\bq_2 \equiv \zhat\cdot(\bq_1\times\bq_2)$.
This algebra is the quantum version of the algebra of
area-preserving diffeomorphisms on the plane and can also be
interpreted as that of magnetic translations in a uniform field~\cite{kogan1994area}.
Together, the density algebra of Eq.~\eqref{GMP} and  the effective Hamiltonian
of Eq.~\eqref{LLL-proj-int} capture the non-trivial dynamics that arise from
projection to the lowest Landau level.  

Since the fractionally filled Chern insulator has an effective
Hamiltonian which has the same form as that of the FQHE, if the
projected density operators of the fractional Chern insulator
also obey the same algebra, the same low energy physics would
ensue, explaining the existence of topological phases in
flattened Chern bands.

Let us therefore examine the projected density operators of the
Chern band, following a strategy  outlined in Ref.~\onlinecite{parameswaran2012fractional}. Let $P_{\alpha} =
\sum_{\k} \ket{\k,\alpha}\bra{\k,\alpha}$ be the operator that
projects to the Chern band. A Taylor expansion of the projected
density operator, $\bar{\rho}_{\q}=P_{\alpha}
\rho_{\q}P_{\alpha}$ keeping only terms of order $q^2$ yields:
\begin{align}
 \bar{\rho}_{\q} = P_{\alpha} + i
P_{\alpha}\q.\r P_{\alpha}  -\frac{1}{2} P_{\alpha}(\q.\r)^2 P_{\alpha}  
\end{align}

From this expression, it follows  that, provided the Berry curvature, $B_{\alpha}(\k)$ is
uniform in momentum space, i.e., provided the fluctuations in
the Berry curvature can be neglected, up to order $q^2$, the following relation holds:
\begin{align}
  [\bar{\rho}_{\q_1},\bar{\rho}_{\q_2}]= i(\q_1 \wedge \q_2) \bar{B}_{\alpha}P_{\alpha} 
\end{align}
where $\bar{B}_{\alpha}= 2\pi C_{\alpha}/A_{BZ}$ is the average
Berry curvature and $A_{BZ}$ is the area of the Brillouin
zone. One may then assert~\cite{parameswaran2012fractional} (that to order
$q^2$), this has the same form as the $W_{\infty}$ algebra of
the LLL projected density
operators
with $\sqrt{\bar{B}}_{\alpha}$ playing the role of the magnetic
length, $l_B$. 
 
Most band structures do not have a uniform Berry curvature, and
thus the relation holds only approximately, even to order
$q^2$. One can however make a virtue of what seems like a
failing by arguing that the degree of deviation from a uniform
Berry curvature provides a way to predict quantify measure how
good a host a particular band structure is for hosting
FQHE-like phases, an expectation that has been confirmed by
numerics~\cite{PhysRevB.85.075116}.
 
 It is natural to consider higher order terms in $q$ in $
 [\bar{\rho}_{\q_1},\bar{\rho}_{\q_2}]$. Keeping terms of order 
$q^3$, we find, after a little algebra, that 
\begin{align*}
   [\bar{\rho}_{\q_1},\bar{\rho}_{\q_2}] = (i)\bar{B}_{\alpha}(\q_1 \wedge \q_2)P_{\alpha}(1 + i
 P_{\alpha}(\q_1+\q_2).\r P_{\alpha}) - \\
\frac{i}{2}\sum_{a,b,c}\left(\frac{q_{1a}q_{2b}q_{2c}}{2}[P_{\alpha}r_{a}P_{\alpha},P_{\alpha}(r_bQ_{\alpha}r_c+
  r_cQ_{\alpha}r_b)P_{\alpha}] \right.
  +\\  \left. 
  \frac{q_{1a}q_{1b}q_{2c}}{2}[P_{\alpha}(r_aQ_{\alpha}r_b+ r_bQ_{\alpha}r_a)P_{\alpha},P_{\alpha}r_cP_{\alpha}]\right)\\
\end{align*}
where $Q_{\alpha}=I - P_{\alpha}$ and $I$ is the identity
operator. 
 The commutators $[P_{\alpha}r_{a}P_{\alpha},P_{\alpha}(r_bQ_{\alpha}r_c+
  r_cQ_{\alpha}r_b)P_{\alpha}]$ and $[P_{\alpha}(r_aQ_{\alpha}r_b+ r_bQ_{\alpha}r_a)P_{\alpha},P_{\alpha}r_cP_{\alpha}]$
vanish if and only if the Fubini-Study (FS) metric tensor,
$g^{\alpha}(\k)$ is a constant in the Brillouin zone. The Fubini-Study metric is a rank two symmetric tensor, $g^{\alpha}(\k)$ with components~\cite{page1987geometrical,PhysRevLett.65.1697,kobayashi1969foundations,Pati1991105} 
\begin{align*}
g^{\alpha}_{ab}(\k)&=\frac{1}{2}\sum_{p}\left[\left(\frac{\partial {u}^{\alpha*}_p}
{\partial k_a}
\frac{\partial{u}^{\alpha}_p}{\partial k_b} + 
\frac{\partial {u}^{\alpha*}_p}{\partial k_b}
\frac{\partial{u}^{\alpha}_p}{\partial k_a} \right)  \right.\\
&\;\;\;\; \left. - \sum_{q}\left(\frac{\partial {u}^{\alpha*}_p}{\partial k_b}u^{\alpha}_p{u}^{\alpha*}_q 
\frac{\partial {u}^{\alpha}_q}{\partial k_a} +  
\frac{\partial {u}^{\alpha*}_p}{\partial k_a}u^{\alpha}_p{u}^{\alpha*}_q \frac{\partial 
{u}^{\alpha}_q}{\partial k_b} \right) \right] \\
\end{align*}

 If the metric tensor is constant in the Brillouin zone, then
 to order $q^3$, the Chern band projected densities satisfy the
 $W_{\infty}$ algebra of projected LLL densities. This leads us
 to identify the uniformity of the metric tensor in momentum
 space as an additional ``metric'' for identifying ``good'' band structures
 from the point of view of hosting interacting topological
 phases. Of course, other conditions such as a suitable short ranged
 interaction, and a proper hierarchy of energy scales are no
 less important. 

 We will see that when the band structure satisfies one
 additional constraint, the Chern band projected densities
 satisfy the $W_{\infty}$ algebra of projected LLL densities at
 all orders in $q$. If one could completely ignore the
 Fubini-Study  metric tensor, i.e., set it to zero, then with the
 assumption of a constant Berry curvature, the algebra of
 projected density operators would simply be the Heisenberg
 algebra and would close at all wavelengths. However, the FS metric cannot vanish as 
the non-trivial topology of the band
 structure of a Chern band places some constraints on the form
 of the FS metric tensor. The trace of the FS metric tensor at
 a given point in $k$ space, which we denote by $tr(g^{\alpha}(\k))$ can
 be expressed as:
 \begin{align}
   tr(g^{\alpha}(\k)&=
   \bra{\k,\alpha}(x+iy)(I-P_{\alpha})(x-iy)\ket{\k,\alpha}
   \nonumber \\ 
                &+ \,i\bra{\k,\alpha}(x(I-P_{\alpha})y-
                y(I-P_{\alpha})x)\ket{\k,\alpha}  \label{metric-trace}
 \end{align}

The positive definiteness of the
 operators, $A^{\dagger}A$ and $C^{\dagger}C$ where
 $A=(I-P_{\alpha})(x+iy)P_{\alpha}$ and
 $C=(I-P_{\alpha})(x-iy)P_{\alpha}$ implies that 
 \begin{align}
  \bra{\k,\alpha}AA^{\dagger}\ket{\k,\alpha} \ge 0 \;\;,\;\; 
  \bra{\k,\alpha}CC^{\dagger}\ket{\k,\alpha} \ge 0 \label{inequalities}
 \end{align}
From these inequalities and Eq.~\eqref{metric-trace}, it follows that 
\begin{align}
  tr(g^{\alpha}(\k)) \ge |B_{\alpha}(\k)| \label{metric-berry-inequality}
\end{align}
 Thus the magnitude of the Berry curvature places a lower bound on the trace of the
 Fubini-Study metric. 

 One may define the transformed operators, $x'_a = \sum_bt_{ab} x_b$,
 corresponding to rotations and scale transformations. Here $t$
 is an invertible matrix and $x_1=x,x_2=y$.

We may also define a corresponding transformed FS metric, $g^{\alpha'}(\k)$:
\begin{align*} 
g^{\alpha'} _{ab}(\k) = \frac{1}{2}\bra{\k,\alpha}\left(x'_a (I-P_{\alpha}) x'_b + x'_b (I-P_{\alpha}) x'_a \right) \ket{\k,\alpha}
\end{align*}
Thus,  $ g^{\alpha'}_{ab}(\k)= t_{ac}g^{\alpha}_{cd}(\k)t_{bd}$. Similarly, the transformed Berry curvature is 
\begin{align*}
B'_{\alpha}(\k) = \sum_{a,b}\epsilon_{ab}\bra{\k,\alpha} x'_a P_{\alpha} x'_b \ket{\k',\alpha}
\end{align*} 

 If one chooses $t$ such that it corresponds to a unimodular
 coordinate transformation with $\det(t)=1$, the Berry curvature
 is left unchanged, i.e., $B'_{\alpha}(\k)=B_{\alpha}(\k)$. The
 inequality~\eqref{metric-berry-inequality} then also applies
 to the transformed FS metric and the Berry curvatures. Thus, 
 \begin{align}
tr(g^{\alpha'}(\k)) \ge |B_{\alpha}(\k)| \label{metric-ineq2} 
 \end{align}

 One can always find a unimodular transformation such that the
 transformed metric at any given point, $\k_0$ is a
 diagonal matrix. Since the determinant of the FS metric is
 preserved through such a transformation, the transformed
 metric may be written as 
\begin{align}
g^{\alpha'}(\k_0) = \begin{pmatrix}
  \sqrt{\det(g^{\alpha}(\k_0))} & 0 \\ 0 &
  \sqrt{\det(g^{\alpha}(\k_0))} \end{pmatrix} \label{diag-metric} 
\end{align}

 Applying the inequality \eqref{metric-ineq2} to the
 transformed metric of Eq.~\eqref{diag-metric}, we find that
$2 \sqrt{\det(g^{\alpha}(\k_0)}\ge |B_{\alpha}(\k_0)|$. Since, the point
$\k_0$ is arbitrary, we conclude that  
\begin{align}
 \det(g^{\alpha}(\k))\ge \frac{|B_{\alpha}(\k)|^2}{4} \label{metric-ineq3}
\end{align}
for any $\k$ in the BZ. 
Further, since
\begin{align} 
 \int_{BZ} dk_x dk_y\, (B_{\alpha}(\k))^2
 \ge A_{BZ}{\bar{B}_{\alpha}}^2\,,
 \end{align} it follows that
\begin{align}
\int_{BZ} dk_x dk_y \, \det(g^{\alpha}(\k))\ge
\frac{A_{BZ}\bar{B}_{\alpha}^2}{4}=
\frac{\pi^2C^2_{\alpha}}{A_{BZ}} \label{final-ineq}
\end{align}
Thus the integral of the determinant of the FS metric is
bounded from below by a number which is proportional to the square of
the topological invariant of the band.

Consider now, the case when the inequality~\eqref{final-ineq}
is saturated and the FS metric is uniform in the BZ. The
inequality~\eqref{final-ineq} is saturated when
$\det(g^{\alpha}(\k)) = \frac{|B_{\alpha}(\k)|^2}{4}$ at all points,
$\k$ in the BZ, and when in addition, the Berry curvature is
uniform in the BZ. From the constancy of the FS metric and the
saturation of inequality~\eqref{metric-ineq3}, it follows that
there is some matrix $t'$ such that
\begin{align}
t' g^{\alpha}(\k) (t')^T = 
\begin{pmatrix} 
\frac{\bar{B}_{\alpha}}{2} & 0 \\
0 & \frac{\bar{B}_{\alpha}}{2}
\end{pmatrix},
\end{align}
where we have assumed without loss of generality that
$\bar{B}_{\alpha}>0$. If $x',y' $ are the corresponding transformed
position operators, from Eq.~\ref{metric-trace} and the above conditions, it follows
that 
\begin{align*}
 \sum_{\k,\gamma}\bra{\k,\gamma}P_{\alpha} (x'+iy')Q_{\alpha}(x' - iy')
P_{\alpha} \ket{\k,\gamma} = 0. 
\end{align*}
This implies that the trace of $DD^{\dagger}$ where $D=
Q_{\alpha}(x'- i y')P_{\alpha}$ is zero and since
$DD^{\dagger}$ is a positive definite matrix,  we may conclude that $
Q_{\alpha}(x' - i y')P_{\alpha}= 0$. 

Let $q'_a =
\sum_{b}t^{-1}_{ab}q_b$. Writing the density operator $\rho_{\q}$
as  $e^{i\q.\r} =e^{i\q'.\r'} =e^{\frac{i}{2}\left\{(q'_x+ i q'_y)(x' - i y') + (q'_x-i q'_y)(x' + i y') \right\}} $, it is easy to verify that the
density operators satisfy a generalized metric-dependent version of the $W_{\infty}$ algebra:
\begin{align*}
[\bar{\rho}_{\q_1},\bar{\rho}_{\q_2}]= 2i \sin(\frac{\q_1 \wedge \q_2 \bar{B}_{\alpha}}{2})e^{(q_1)_l g^{\alpha}_{lm}(q_2)_m}\bar{\rho}_{\q_1 +
  \q_2}
\end{align*}
In summary, if $\int_{BZ} dk_x dk_y \,
\det(g^{\alpha}(\k))=\frac{\pi^2C^2_{\alpha}}{A_{BZ}}$ and if
the Fubini-Study metric is uniform in the BZ, the density
operators satisfy a closed algebra which is a generalization of
the usual $W_{\infty}$ algebra. We note a couple of interesting
features. Firstly, the Berry curvature and the FS metric both
appear in this form of the $W_{\infty}$ algebra. Thus the
algebra also applies to bands, which have a higher Chern number
and which therefore differ fundamentally from Landau levels
which have Chern number 1. Secondly, we observe that the
conditions under which we get a closed algebra of the projected
density operators can be stated purely in terms of the FS
metric.

For a system where the ideal conditions under which this
algebra is obtained do not hold, the degree of deviation from
these conditions provides a new parameter (or a set of
parameters, depending on how one chooses to quantify the
deviation) to predict how favorable a Chern band is for hosting
FQHE-like physics. Conversely, if one finds fractional
topological phases in systems where the deviations from these
conditions is considerable, one could argue that the physics of
those systems is new and different from the conventional
fractional quantum Hall effect.

The effects of disorder also enter the Hamiltonian through
terms that involve the projected density operator. This
suggests that the effects of disorder in the Chern band are
likely to be the same as in the LLL when the conditions stated
above for the FS metric are satisfied.

Let us briefly discuss other band structures where fractional
phases may
arise~\cite{2009PhRvL.103s6803L,2011PhRvB..84w5145L,PhysRevB.84.165107}. One
primary example are topological bands with time-reversal
symmetry. Consider, for instance, the case of $Z_2$ insulators
with a pair of time-reversed paired flat bands.  The projection
operator, $P$ to the topological bands can then always be
written as a sum $P = P_{1}+ P_{2}$ where $P_1$ and $P_2$ are a
pair of projectors related by time-reversal symmetry, which
have Chern numbers associated with them that are equal in
magnitude and opposite in
sign~\cite{Roy_09_195321,roy2010characterization}. There may be
circumstances where the interactions between electrons with
different indices can be neglected either due to the nature of
the physical interactions or due to the formation of
fractionalized states where the role of such interactions are
minimized. Then the relevant projected density operators are
$P_{i}\rho_{\q}P_{i}$ ($i=1,2$) and the conditions under
which these form a closed algebra are the uniformity of the FS
metric associated with each projection operator and the
saturation of inequality~\eqref{final-ineq} for the same.

Ref.~\cite{parameswaran2012fractional} and the current work highlight the
important role of geometric features of bands in fractional
topological insulators. One is tempted to even go so far as to
suggest that the term ``fractional topological insulators"
should be replaced by ``fractional geometric insulators".

I thank Siddharth Paramewsaran and Shivaji Sondhi for
discussions, comments and an earlier
collaboration. Additionally, I would like to thank Sudip
Chakarvarty, John Chalker, Steve Simon and R Shankar for
helpful conversations and comments.

\end{document}